\documentclass[aps,preprintnumbers,epsf,amssymb,graphicx]{revtex4}
\usepackage[T1]{fontenc}           
\usepackage[latin1]{inputenc}      

\begin{document}

\title{Non-perturbative scalar gauge-invariant metric fluctuations from the Ponce de Le\'on metric in the
STM theory of gravity}
\author{$^{1,2}$Mariano Anabitarte \footnote{
E-mail address: anabitar@mdp.edu.ar}, $^{1,2}$Mauricio
Bellini\footnote{E-mail address: mbellini@mdp.edu.ar}}

\address{$^1$ Departamento de
F\'{\i}sica, Facultad de Ciencias Exactas y Naturales, Universidad
Nacional de Mar del Plata, Funes 3350, (7600) Mar del Plata,
Argentina.\\
$^2$ Consejo Nacional de Investigaciones Cient\'{\i}ficas y
T\'ecnicas (CONICET).}

\begin{abstract}
We study our non-perturbative formalism to describe scalar
gauge-invariant metric fluctuations by extending the Ponce de
Le\'on metric.
\end{abstract}
\maketitle

\section{Introduction}

The possibility that our world may be embedded in a
$(4+d)$-dimensional universe with more than four large dimensions
has attracted the attention of a great number of researches. One
of these higher-dimensional theories, where the cylinder condition
of the Kaluza-Klein theory\cite{kk} is replaced by the conjecture
that the ordinary matter and fields are confined to a 4D subspace
usually referred to as a brane is the Randall and Sundrum
model\cite{rs}.

Another non-compact theory is the so called Space - Time - Matter
(STM) or Induced Matter (IM) theory. In this theory the conjecture
is that the ordinary matter and fields that we observe in 4D
result from the geometry of the extra dimension\cite{im}. In this
framework, inflationary models induced from a 5D vacuum state,
where the expansion of the universe is driven by a single scalar
(inflaton) field, has been subject of great activity in the last
years\cite{nuestros}. The scalar metric fluctuations related to
the inflaton field fluctuations can be studied as invariant under
gauge transformations in a standard 4D cosmological
model\cite{brandenberger}, or from a 5D vacuum theory of
gravity\cite{nu}. These perturbations are related with energy
density perturbations. They are spin-zero projections of the
graviton, which only exist in non-vacuum cosmologies. The issue of
gauge invariance becomes critical when we attempt to analyze how
the scalar metric fluctuations $\psi$ produced in the very early
universe influence the expansion with respect to the 4D background
isotropic, homogeneous and 3D spatially flat cosmological metric.
From the cosmological point of view, these metric fluctuations are
produced by the inflaton field fluctuations $\varphi
-\left<\varphi\right>$, which describe the quantum fluctuations of
the inflaton field with respect to the expectation value of this
field on the 3D sphere: $\left<\varphi\right>$, in the absence of
metric fluctuations. In other words, from the relativistic point
of view, the quantum field fluctuations of the inflaton field
(which is a scalar field) induce the quantum (scalar) metric
fluctuations on the background metric. From the mathematical point
of view the metric fluctuations are the geometrical deformations
produced by quantum field fluctuations of the inflaton field.

We consider the 5D background line element\cite{pdl}
\begin{equation}\label{bam}
dS^2 =  l^2 dt^2 - \left( \frac{t}{t_0} \right) ^{2p}
l^{\frac{2p}{p-1}} dr^2 - \frac{t^2}{\left( p-1 \right)^2} dl^2,
\end{equation}
were $dr^2=dx^2+dy^2+dz^2$, and $l$ is the non-compact extra
dimension and $p$ is a dimensionless constant. This metric is 3D
spatially isotropic, homogeneous and Riemann flat:
$\bar{R}_{ABCD}=0$ ($A$ and $B$ run from $0$ to $4$), but it is
curved in four dimensions\cite{seahra}. From the physical point of
view, this metric represents an apparent vacuum $\bar{G}_{AB}=0$,
which deserves interest in space-time-matter theory. The
particular case in which we take a foliation $l=l_0$ it is very
important for cosmology\cite{pdl,librowesson}
\begin{equation}\label{4b}
dS^2 = l^2_0 dt^2 - \left(\frac{t}{t_0}\right)^{2p}
l^{\frac{2p}{p-1}}_0 dr^2,
\end{equation}
because it describes an effective 4D universe that expands with a
scale factor $a(t) \sim t^p$, with a pressure ${\rm
P}={(2-3p)p\over 8\pi G l^2_0 t^2}$ and an energy density $\rho=
{3 p^2\over 8\pi G l^2_0 t^2}$. In particular, in the limit case
in which $p \rightarrow \infty$, the metric (\ref{4b}) describes
an inflationary expansion of the universe\cite{ab} with a vacuum
dominated equation of state ${\rm P} \simeq -\rho$. Other
important cases are $p=1/2,2/3$, which describe, respectively,
radiation and matter dominated universes in absence of vacuum. In
this letter we shall study a non-perturbative formalism for
gauge-invariant metric fluctuations $\psi(x)$ in the STM theory of
gravity, starting with the Ponce de Le\'on
metric (\ref{bam}).\\

\section{Formalism}

With the aim to study strong gauge-invariant (scalar) metric
fluctuations, we propose the following metric:
\begin{equation}\label{met}
dS^2 =  l^2 e^{2\psi} dt^2 - \left( \frac{t}{t_0} \right) ^{2p}
l^{\frac{2p}{p-1}} e^{-2\psi} dr^2 - \frac{t^2}{\left( p-1
\right)^2} e^{-2\psi} dl^2,
\end{equation}
where $\psi(t,x,y,z,l)$ is a quantum scalar field. This metric is
a generalization for strong gauge-invariant (scalar) metric
fluctuations of one previously studied in \cite{ab1}, which is
only valid for small metric fluctuations: $e^{\pm 2\psi}\simeq
1\pm 2\psi $. To describe the system in an apparent vacuum, we
shall consider the action
\begin{equation}
^{(5)}I = {\Large\int} d^4 x \  dl \sqrt{\left|\frac{^{(5)}
 g}{^{(5)} g_0}\right|} \left(
\frac{^{(5)} R}{16\pi G}+ \frac{1}{2}g^{AB} \varphi_{,A}
\varphi_{,B} \right),
\end{equation}
where $^{(5)}g$ is the determinant of the covariant metric tensor
$g_{AB}$:
\begin{equation}\label{ricci5}
^{(5)}g=\left[t \left(\frac{t}{t_0}\right)^{3p}
l^{\frac{4p-1}{p-1}} \frac{e^{-3\psi}}{(p-1)}\right]^2.
\end{equation}

\subsection{Lagrange equations in a 5D apparent vacuum}

The Ricci Scalar, which in our case is null, being given by the
expression
\begin{eqnarray}
^{(5)} R & = & 4 \; l^{\frac{-2p}{p-1}}
\left(\frac{t}{t_0}\right)^{-2p} e^{2\psi}
\left\{\nabla^2\psi-(\nabla\psi)^2+ e^{-4\psi} l^{\frac{2}{p-1}}
\left(\frac{t}{t_0}\right)^{2p}\left[7 (\psi_{,t})^2 - 2\psi_{,tt}
- \frac{3}{t} (3p+1) \psi_{,t}\right] \right.\nonumber \\
&& \left. + \frac{l^{\frac{2p}{p-1}}}{t^2} \left( \frac{t}{t_0}
\right)^{2p} (p-1) \left[ (p-1) \psi_{,ll} - (p-1) (\psi_{,l})^2 +
\frac{3p}{l} \psi_{,l} \right] - \frac{3 p^2}{t^2} \;
l^{\frac{2}{p-1}} \left( \frac{t}{t_0} \right)^{2p} \left[ 1 -
e^{-4 \psi} \right] \right\},
\end{eqnarray}
where $\psi_{,A} = \frac{\partial}{\partial A}$. The Lagrange
equations give us the relevant equations of motion for the fields
$\varphi$ and $\psi$, respectively:
\begin{eqnarray}
\varphi_{,tt} & + & \left[ \frac{(3p+1)}{t} - 5 \psi_{,t}\right]
\varphi_{,t} -  e^{4\psi} \left(\frac{t}{t_0}\right)^{-2p}
l^{\frac{-2}{p-1}} \left( \nabla^2 \varphi - \vec\nabla \psi .
\vec\nabla \varphi \right) \nonumber \\
& - & \frac{l^2}{t^2} e^{4\psi} (p-1)^2 \left[ \varphi_{,ll} +
\left( \frac{(4p-1)}{(p-1)} l^{-1} - \psi_{,l} \right)
\varphi_{,l} \right]=0, \label{lag1} \\
\left( \frac{\partial^{(5)}R}{\partial\psi} - 3 \,{^{(5)} R}
\right) & - & \left[ \frac{1}{\sqrt{\left|^{(5)}g\right|}}
\frac{\partial\sqrt{\left|^{(5)}g\right|}}{\partial x^A}
\frac{\partial^{(5)}R}{\partial \psi_{,A}} +
\frac{\partial}{\partial x^{A}} \left(
\frac{\partial^{(5)}R}{\partial \psi_{,A}}\right) \right]
\nonumber \\
& = & 8\pi G \left[ 5 l^{-2} e^{-2 \psi} (\varphi_{,t})^2
-\left(\frac{t}{t_0}\right)^{-2p} l^{\frac{-2p}{p-1}} e^{2 \psi}
(\nabla \varphi)^2 - e^{2\psi} \frac{(p-1)^2}{t^2}(\varphi_{,l})^2
\right]. \label{lag2}
\end{eqnarray}
The 5D energy momentum tensor for a scalar field $\varphi$ in the
absence of interactions, is
\begin{equation}\label{mom}
T_{AB} = \varphi_{,A} \varphi_{,B} - \frac{1}{2} g_{AB}
\varphi_{,C} \varphi^{,C},
\end{equation}
where the components $g_{AB}$ are given by the perturbed metric
(\ref{met}). Using the fact that $T^t_{\,\,t}$ is given by the
expression
\begin{equation}
T^t_{\,\,t} = \frac{1}{2} \left[l^{-2} e^{-2 \psi}
(\varphi_{,t})^2+ \left(\frac{t}{t_0}\right)^{-2p}
l^{\frac{-2p}{p-1}} e^{2 \psi} (\nabla \varphi)^2+ e^{2\psi}
\frac{(p-1)^2}{t^2}(\varphi_{,l})^2\right],
\end{equation}
(\ref{lag2}) can be written in a more compact manner as
\begin{equation}
\left( \frac{\partial^{(5)}R}{\partial\psi} - 3 \,{^{(5)} R}
\right) -  \left[ \frac{1}{\sqrt{\left|^{(5)}g\right|}}
\frac{\partial\sqrt{\left|^{(5)}g\right|}}{\partial x^A}
\frac{\partial^{(5)}R}{\partial \psi_{,A}} +
\frac{\partial}{\partial x^{A}} \left(
\frac{\partial^{(5)}R}{\partial \psi_{,A}}\right) \right] =8 \pi G
\left[ 6 l^{-2} e^{-2 \psi} (\varphi_{,t})^2 - 2 T^t_{\,\,t}
\right]. \label{lag3}
\end{equation}
Equations (\ref{lag1}) and (\ref{lag3}) relate the quantum field
$\varphi$ with (quantum gauge-invariant) scalar metric
fluctuations.

\subsection{5D Einstein equations on an apparent vacuum}

To obtain the Einstein equations, we shall calculate the
components of the Einstein tensor. Their diagonal components (we
consider $G_{rr}=G_{xx}+G_{yy}+G_{zz}$) are given by
\begin{eqnarray}
G_{tt} =&& \frac{- 3}{t^2} \left\{ 2 \; t^2 (\psi_{,t} )^2 -
(3p+1) \; t \; \psi_{,t} -  e^{4\psi} \; l^{\frac{-2}{p-1}}
\left(\frac{t}{t_0}\right)^{-2p} t^2 \left[ (\nabla \psi)^2 -
\nabla^2 \psi \right] \right.\nonumber \\
&&\left. + \; e^{4\psi} \; l^2 (p-1)^2 \left[ \psi_{,ll} -
(\psi_{,l})^2 - \frac{3p}{l (p-1)} \psi_{,l}\right] - p(p-1)\left[
e^{4\psi} -1 \right] \right\}, \label{G1}
\end{eqnarray}

\begin{eqnarray}
G_{rr} =&&\frac{-3}{l^2 t^2} \left\{ e^{-4\psi} l^{\frac{2p}{p-1}}
\left(\frac{t}{t_0}\right)^{2p} \left[ 3 t^2 \psi_{,tt} - 9 t^2
(\psi_{,t})^2 + 5 (2p+1) t \psi_{,t} \right] + \frac{2}{3} l^2 t^2
\left[ (\nabla \psi)^2 - \nabla^2 \psi \right] \right.\nonumber \\
&& \left. + \left(\frac{t}{t_0}\right)^{2p}
l^{\frac{2(p-1)}{p-1}}(p-1)^2 \left[ (\psi_{,l})^2 - \psi_{,ll} -
\frac{(p^2-1)}{l(p-1)^2} \psi_{,l} \right] + 3
\left(\frac{t}{t_0}\right)^{2p} l^{\frac{2p-1}{p-1}} p^2 \left[ 1-
e^{-4\psi} \right] \right\}, \label{G2}
\end{eqnarray}

\begin{eqnarray}
G_{ll} =&& - \frac{1}{l^2 (p-1)^2} \left\{ e^{-4\psi} \left[ 3 t^2
\psi_{,tt} - 9 t^2 (\psi_{,t})^2 + 15 p t \psi_{,t} \right] +
l^{\frac{-2}{p-1}} \left(\frac{t}{t_0}\right)^{-2p} t^2 \left[
(\nabla \psi)^2 - \nabla^2 \psi \right] \right.\nonumber \\
&& \left. + l (-6p^2+9p-3) \psi_{,l} + 3 p (2p-1) \left[ 1 -
e^{-4\psi} \right] \right\}. \label{G3}
\end{eqnarray}

Since we require the Ricci scalar (\ref{ricci5}) to be null:
$^{(5)} R=0$, we obtain

\begin{eqnarray}
\left[ 1 - e^{-4 \psi} \right] = && \frac{t^2}{3 p^2} \;
l^{\frac{-2}{p-1}} \left( \frac{t}{t_0} \right)^{-2p} \left\{
\nabla^2 \psi - (\nabla \psi)^2 + e^{-4\psi} l^{\frac{2}{p-1}}
\left(\frac{t}{t_0}\right)^{2p}\left[7 (\psi_{,t})^2 - 2\psi_{,tt}
- \frac{3}{t} (3p+1) \psi_{,t}\right] \right.\nonumber \\
&& \left. + \frac{l^{\frac{2p}{p-1}}}{t^2} \left( \frac{t}{t_0}
\right)^{2p} (p-1) \left[ (p-1) (\psi_{,ll} - (\psi_{,l})^2 ) +
\frac{3p}{l} \psi_{,l} \right] \right\}, \label{equa}
\end{eqnarray}
so that, using (\ref{equa}) in (\ref{G1})-(\ref{G3}), we obtain
the 5D diagonal components of the Einstein tensor:
\begin{eqnarray}
G_{tt} =&& \frac{- 3}{t^2} \left\{ - t^2 \frac{(p+7)}{3p}
(\psi_{,t} )^2 + t^2 \frac{2(p+1)}{3p} \psi_{,tt} +
\frac{(3p+1)}{p} \; t \; \psi_{,t} + e^{4\psi} \;
l^{\frac{-2}{p-1}} \left(\frac{t}{t_0}\right)^{-2p} t^2 \left[
(\nabla \psi)^2 -
\nabla^2 \psi \right] \right.\nonumber \\
&&\left. + \; e^{4\psi} \; l^2 (p-1)^2 \frac{2p-1}{3p} \left[
\psi_{,ll} - (\psi_{,l})^2 - \frac{3p}{l (2p-1)}
\frac{(4p+1)}{(p-1)} \psi_{,l}\right]\right\},
\end{eqnarray}

\begin{eqnarray}
G_{rr} =&&\frac{-3}{l^2 t^2} \left\{ e^{-4\psi} l^{\frac{2p}{p-1}}
\left(\frac{t}{t_0}\right)^{2p} \left[ t^2 \psi_{,tt} - 2 t^2
(\psi_{,t})^2 + (p+2) t \psi_{,t} \right] - \frac{l^2 t^2}{3}
\left[ (\nabla \psi)^2 - \nabla^2 \psi \right] \right.\nonumber \\
&& \left. + \left(\frac{t}{t_0}\right)^{2p}
l^{\frac{3p-1}{p-1}}(2p^2-3p+1) \psi_{,l} \right\}
\end{eqnarray}

\begin{eqnarray}
G_{ll} =&& - \frac{1}{l^2 (p-1)^2} \left\{ e^{-4\psi} \left[
\frac{(2-p)}{p} t^2 \psi_{,tt} + \frac{(5p-7)}{p} t^2
(\psi_{,t})^2 - 3 \frac{\left( p^2-p-1\right)}{p} t \psi_{,t} \right] \right.\nonumber \\
&& \left. - \frac{(p-1)}{p} l^{\frac{-2}{p-1}}
\left(\frac{t}{t_0}\right)^{-2p} t^2 \left[ (\nabla \psi)^2 -
\nabla^2 \psi \right] + l^2 \frac{(2p-1)}{p} (p-1)^2 \left[
\psi_{,ll} - (\psi_{,l})^2 \right] \right\}.
\end{eqnarray}

The diagonal components $T_{AB}$ of the (covariant) energy
momentum tensor are [see the expression (\ref{mom})]

\begin{eqnarray}
T_{tt} & = & \frac{1}{2} \left\{(\varphi_{,t})^2 + e^{4\psi}
l^{\frac{-2}{p-1}} \left(\frac{t}{t_0}\right)^{-2p} (\nabla
\varphi)^2 +\frac{l^2}{t^2} e^{4\psi} (p-1)^2
(\varphi_{,l})^2\right\}, \label{t1} \\
T_{ll} & = &  \frac{1}{2} \left\{ (\varphi_{,l})^2 -
\frac{t^2}{l^2} \frac{e^{-4\psi}}{(p-1)^2} (\varphi_{,t})^2 +
\frac{t^2}{(p-1)^2} l^{\frac{-2p}{p-1}}
\left(\frac{t}{t_0}\right)^{-2p} (\nabla
\varphi)^2 \right\}, \label{t2} \\
T_{xx} & = &  (\varphi_{,x})^2 + \frac{1}{2} l^{\frac{2}{p-1}}
\left(\frac{t}{t_0}\right)^{2p} e^{-4\psi} (\varphi_{,t})^2 -
\frac{1}{2} (\nabla \varphi)^2 - \frac{1}{2}
l^{\frac{2p}{p-1}} \left(\frac{t}{t_0}\right)^{2p} \frac{(p-1)^2}{t^2}
(\varphi_{,l})^2, \label{t3} \\
T_{yy} & = & (\varphi_{,y})^2 + \frac{1}{2} l^{\frac{2}{p-1}}
\left(\frac{t}{t_0}\right)^{2p} e^{-4\psi} (\varphi_{,t})^2 -
\frac{1}{2} (\nabla \varphi)^2 - \frac{1}{2} l^{\frac{2p}{p-1}}
\left(\frac{t}{t_0}\right)^{2p} \frac{(p-1)^2}{t^2}
(\varphi_{,l})^2, \label{t4} \\
T_{zz} & = & (\varphi_{,z})^2 + \frac{1}{2} l^{\frac{2}{p-1}}
\left(\frac{t}{t_0}\right)^{2p} e^{-4\psi} (\varphi_{,t})^2 -
\frac{1}{2} (\nabla \varphi)^2 - \frac{1}{2} l^{\frac{2p}{p-1}}
\left(\frac{t}{t_0}\right)^{2p} \frac{(p-1)^2}{t^2}
(\varphi_{,l})^2. \label{t5}
\end{eqnarray}
Since the metric (\ref{met}) is 3D spatially  isotropic, we can
make the identification $T_{rr} = T_{xx} + T_{yy} + T_{zz}$, and
we obtain

\begin{equation}\label{T2}
T_{rr}  = \frac{3}{2} l^{\frac{2}{p-1}}
\left(\frac{t}{t_0}\right)^{2p} e^{-4\psi} (\varphi_{,t})^2 -
\frac{1}{2} (\nabla \varphi)^2  - \frac{3}{2} l^{\frac{2p}{p-1}}
\left(\frac{t}{t_0}\right)^{2p} \frac{(p-1)^2}{t^2}
(\varphi_{,l})^2.
\end{equation}

Finally, using the expression (\ref{G1})-(\ref{G3}) with
(\ref{t1}), (\ref{T2}) and (\ref{t2}), we obtain

\begin{eqnarray}
&& \left\{ \frac{2(p+1)}{3p} \psi_{,tt} - \frac{(p+7)}{3p}
(\psi_{,t} )^2 + \frac{(3p+1)}{p} \; \frac{1}{t} \; \psi_{,t} +
e^{4\psi} \; l^{\frac{-2}{p-1}} \left(\frac{t}{t_0}\right)^{-2p}
\frac{1-2p}{3p}\left[ (\nabla \psi)^2 - \nabla^2 \psi \right]
\right.
\nonumber \\
&& + \left. \; e^{4\psi} \; \frac{l^2}{t^2} (p-1)^2
\frac{2p-1}{3p} \left[ \psi_{,ll} - (\psi_{,l})^2 - \frac{3p
(4p+1)}{l
(2p-1)(p-1)} \psi_{,l}\right]\right\} \nonumber \\
& &=  \frac{4 \pi G}{3} \left\{(\varphi_{,t})^2  + e^{4\psi}
l^{\frac{-2}{p-1}} \left(\frac{t}{t_0}\right)^{-2p} (\nabla
\varphi)^2 +\frac{l^2}{t^2} e^{4\psi} (p-1)^2
(\varphi_{,l})^2\right\}, \label{e1}
\end{eqnarray}
\begin{eqnarray}
&& e^{-4\psi} l^{\frac{2}{p-1}} \left(\frac{t}{t_0}\right)^{2p}
\left[ \psi_{,tt} - 2 (\psi_{,t})^2 + \frac{(p+2)}{t}
\psi_{,t}\right] - \frac{1}{3} \left[ (\nabla \psi)^2 - \nabla^2
\psi \right] + \left(\frac{t}{t_0}\right)^{2p}
\frac{l^{\frac{p+1}{p-1}}}{t^2}(2p^2-3p+1) \psi_{,l} \nonumber
\\
&& = \frac{4 \pi G}{3} \left\{ 3 l^{\frac{2}{p-1}}
\left(\frac{t}{t_0}\right)^{2p} e^{-4\psi} (\varphi_{,t})^2 -
(\nabla \varphi)^2  - 3 l^{\frac{2p}{p-1}}
\left(\frac{t}{t_0}\right)^{2p} \frac{(p-1)^2}{t^2}
(\varphi_{,l})^2 \right\}, \label{e2}
\end{eqnarray}
\begin{eqnarray}
&& \frac{t^2}{l^2} e^{-4\psi} \left[ \frac{(2-p)}{p} \psi_{,tt} +
\frac{(5p-7)}{p}(\psi_{,t})^2 - \frac{3}{t}
\frac{p^2-p-1}{p}\psi_{,t} \right]  + \frac{(p-1)}{p}
l^{\frac{-2(2p-1)}{p-1}} \left(\frac{t}{t_0}\right)^{-2p} t^2
\left[ (\nabla \psi)^2 -
\nabla^2 \psi \right] \nonumber \\
&& - \frac{(2p-1)(p-1)^2}{p} \left[ \psi_{,ll} - (\psi_{,l})^2
\right] = -\frac{4 \pi G}{3} \left\{ (p-1)^2(\varphi_{,l})^2 -
\frac{t^2}{l^2} e^{-4\psi} (\varphi_{,t})^2 + t^2
l^{\frac{-2p}{p-1}} \left(\frac{t}{t_0}\right)^{-2p} (\nabla
\varphi)^2 \right\}. \label{e3}
\end{eqnarray}
Equations (\ref{e1}), (\ref{e2}) and (\ref{e3}) give us the
diagonal Einstein equations on a 5D apparent vacuum. In the
following section we shall use these equations to describe the
effective 4D physics on a curved 4D hypersurface which is embedded
on the perturbed metric (\ref{met}).

\section{Effective 4D dynamics of $\varphi$ and gauge-invariant
metric fluctuations $\psi$}

In order to study the effective 4D dynamics of the fields
$\varphi$ and $\psi$, we can make a foliation $l=l_0$, so that
$dl=0$ and the perturbed 4D hypersurface of (\ref{4b}), results to
be
\begin{equation}\label{4met}
dS^2 = l^2_0 e^{2\psi} dt^2 - \left(\frac{t}{t_0}\right)^{2p}
l^{\frac{2p}{p-1}}_0 e^{-2\psi} dr^2.
\end{equation}
An interesting limital case of the metric (\ref{4met}) is that in
which $p\rightarrow \infty$. In such a case the metric
(\ref{4met}) describes an asymptotically vacuum expansion (i.e., a
de Sitter expansion), which should be relevant to describe the
early (inflationary) universe. The particular case where the
gauge-invariant metric fluctuations are weak, was studied in
\cite{ab1}. The effective 4D action is $^{(4)} I = \int d^4x \,
^{(4)} L$, where $^{(4)} L$ is the effective 4D Lagrangian
\begin{equation}
^{(4)} L = \left.\sqrt{\left|\frac{^{(4)}
 g}{^{(4)} g_0}\right|} \left( \frac{^{(4)} R}{16\pi G}+
\frac{1}{2}g^{\mu\nu} \varphi_{,\mu} \varphi_{,\nu} - V
\right)\right|_{l=l_0},
\end{equation}
such that $V$ is the effective 4D potential induced by the
foliation $l=l_0$:
\begin{equation}
V = -\left.\frac{1}{2} g^{ll}
\left(\frac{\partial\varphi}{\partial l}\right)^2\right|_{l=l_0},
\end{equation}
and $^{(4)} R$ is the effective 4D Ricci scalar whose origin is
also geometrically induced by the foliation on the flat metric
(\ref{met})
\begin{equation}
^{(4)} R = \frac{2 e^{-2\psi}}{l_0^2 t^2} \left\{ 3p(2p-1) - t^2
\left(\frac{t}{t_0}\right)^{-2p} l^{\frac{-2p}{p-1}}_0 e^{4\psi}
\left[ (\nabla \psi)^2 - \nabla^2 \psi \right] -3 t  \left[5p
\psi_{,t} - 3 t (\psi_{,t})^2 + t \psi_{,tt}\right] \right\}.
\end{equation}

The effective 4D (diagonal) Einstein equations for the components
$tt$, $rr$ and $ll$ are given respectively by
\begin{eqnarray}
&&\left\{ - \frac{(p+7)}{3p} (\psi_{,t} )^2 + \frac{2(p+1)}{3p}
\psi_{,tt} + \frac{(3p+1)}{p} \; \frac{1}{t} \; \psi_{,t} +
e^{4\psi} \; l_0^{\frac{-2}{p-1}} \left(\frac{t}{t_0}\right)^{-2p}
\frac{1-2p}{3p}\left[ (\nabla \psi)^2 -
\nabla^2 \psi \right] \right.\nonumber \\
&& \left.\left. + \; e^{4\psi} \; \frac{l_0^2}{t^2} (p-1)^2
\frac{2p-1}{3p} \left[ \psi_{,ll} - (\psi_{,l})^2 - \frac{3p
(4p+1)}{l_0 (2p-1)(p-1)} \psi_{,l} \right] \right\}
\right|_{l=l_0} \nonumber \\
&& = \left. \frac{4 \pi G}{3} \left\{ (\varphi_{,t})^2 + e^{4\psi}
l_0^{\frac{-2}{p-1}} \left(\frac{t}{t_0}\right)^{-2p} (\nabla
\varphi)^2 +\frac{l_0^2}{t^2} e^{4\psi} (p-1)^2 (\varphi_{,l})^2
\right\}\right|_{l=l_0},
\end{eqnarray}

\begin{eqnarray}
&& \left. e^{-4\psi} l_0^{\frac{2}{p-1}}
\left(\frac{t}{t_0}\right)^{2p} \left[ \psi_{,tt} - 2
(\psi_{,t})^2 + \frac{(p+2)}{t} \psi_{,t}\right] - \frac{1}{3}
\left[ (\nabla \psi)^2 - \nabla^2 \psi \right] +
\left(\frac{t}{t_0}\right)^{2p}
\frac{l_0^{\frac{p+1}{p-1}}}{t^2}(2p^2-3p+1)
\psi_{,l}\right|_{l=l_0} \nonumber\\
&& = \left.\frac{4 \pi G}{3} \left\{ 3 l_0^{\frac{2}{p-1}}
\left(\frac{t}{t_0}\right)^{2p} e^{-4\psi} (\varphi_{,t})^2 -
(\nabla \varphi)^2  - 3 l_0^{\frac{2p}{p-1}}
\left(\frac{t}{t_0}\right)^{2p} \frac{(p-1)^2}{t^2}
(\varphi_{,l})^2 \right\}\right|_{l=l_0},
\end{eqnarray}

\begin{eqnarray}
&&  \frac{t^2}{l_0^2} e^{-4\psi} \left[ \frac{(2-p)}{p} \psi_{,tt}
+ \frac{(5p-7)}{p}(\psi_{,t})^2 - \frac{3}{t}
\frac{p^2-p-1}{p}\psi_{,t} \right]  + \frac{(p-1)}{p}
l_0^{\frac{-2(2p-1)}{p-1}} \left(\frac{t}{t_0}\right)^{-2p} t^2
\left[ (\nabla \psi)^2 -
\nabla^2 \psi \right] \nonumber \\
&& \left.\left. - \frac{(2p-1)(p-1)^2}{p} \left[ \psi_{,ll} -
(\psi_{,l})^2 \right]\right|_{l=l_0} = -\frac{4 \pi G}{3} \left\{
(p-1)^2(\varphi_{,l})^2 - \frac{t^2}{l_0^2} e^{-4\psi}
(\varphi_{,t})^2 + t^2 l_0^{\frac{-2p}{p-1}}
\left(\frac{t}{t_0}\right)^{-2p} (\nabla \varphi)^2 \right\}
\right|_{l=l_0}. \label{e3}
\end{eqnarray}
The effective Lagrangian equations are
\begin{eqnarray}
\varphi_{,tt} & + & \left[ \frac{(3p+1)}{t} - 5 \psi_{,t}\right]
\varphi_{,t} -  e^{4\psi} \left(\frac{t}{t_0}\right)^{-2p}
l_0^{\frac{-2}{p-1}} \left( \nabla^2 \varphi - \vec\nabla \psi .
\vec\nabla \varphi \right) \nonumber \\
& - & \left.\frac{l_0^2}{t^2} e^{4\psi} (p-1)^2 \left[
\varphi_{,ll} + \left( \frac{(4p-1)}{(p-1)} l_0^{-1} - \psi_{,l}
\right)
\varphi_{,l} \right]\right|_{l=l_0} =0. \\
\left( \frac{\partial^{(4)}R}{\partial\psi} - 3 \,{^{(4)} R}
\right) & - & \left.\left[ \frac{1}{\sqrt{\left|^{(4)}g\right|}}
\frac{\partial\sqrt{\left|^{(4)}g\right|}}{\partial x^A}
\frac{\partial^{(4)}R}{\partial \psi_{,A}} +
\frac{\partial}{\partial x^{A}} \left(
\frac{\partial^{(4)}R}{\partial \psi_{,A}}\right)
\right]\right|_{l=l_0}
\nonumber \\
& = & \left. 8\pi G \left[  6 l^{-2} e^{-2 \psi} (\varphi_{,t})^2
- 2 T^t_{\,\,t} \right]\right|_{l=l_0} ,
\end{eqnarray}
which, like the Einstein equations (\ref{e1}), (\ref{e2}) and
(\ref{e3}), are non-linear. They give us the dynamics of the
system described by the fields $\varphi $ and $\psi $.

Since the field $\varphi(t,\vec r)$ is of quantum origin, it
should be described by the following non-commutative algebra
\begin{equation}
\left[\varphi(t,\vec r), \Pi_{\varphi}(t,\vec r')\right] = i \,
g^{tt} \sqrt{\left|\frac{^{(4)}
 g}{^{(4)} g_0}\right|}\,e^{-\int
\left[\frac{3p+1}{t}- 5\psi_{,t}\right] dt} \,\delta^{(3)}\left(\vec r-\vec r'\right), \\
\end{equation}
where $\left|^{(4)} g\right|=\left(e^{-2 \psi}
\left(\frac{t}{t_0}\right)^{3p} l_0^{\frac{4p-1}{p-1}}\right)^2$
is the determinant of the effective 4D perturbed metric tensor
$g_{\mu\nu}$. The canonical momentum $\Pi_{\varphi}={\partial
^{(4)} L \over
\partial\dot\varphi}$ is given by
\begin{equation}
\Pi_{\varphi} = g^{tt} \; \sqrt{\left|\frac{^{(4)}
 g}{^{(4)} g_0}\right|} \, \;\dot\varphi.
\end{equation}
Of course, due to the non-linear nature of the Einstein equations,
it is almost impossible to resolve the field equations without
making some approximation.

\section{Final Comments}

We have extended to the Ponce de Le\'on metric in the
non-perturbative formalism proposed in \cite{ab1}. To do this, we
have introduced the metric (\ref{met}), which, once we take a
foliation $l=l_0$, takes into account the gauge-invariant metric
fluctuations during the expansion of the universe at its origin
[see eq. (\ref{4met})], as a back reaction effect of the inflaton
field fluctuations $\varphi-\left<\varphi\right>$. The background
4D version of the Ponce de Le\'on metric is very important for
cosmology, because it describes a power-law expansion for the
universe. The interesting thing of this formalism is that the
system is considered from the point of view of a 5D perturbed flat
metric, on which we assume an apparent vacuum state. Hence, all 4D
sources come from the geometrical foliation $l=l_0$ on the 5D
metric (\ref{met}) (which is Riemann flat). The advantage of this
formalism with respect to another one previously
introduced\cite{ab} should be in the description of the strong
metric fluctuations, which should be more important in the early
universe on very small scales. Of course, the results obtained
in\cite{ab} should here be recovered in the weak field
approximation. In this approximation back reaction effects become
negligible and the 4D version of the Ponce de Le\'on metric [see
the eq. (\ref{4b})], describes in the limit case $p \rightarrow
\infty$ a vacuum dominated expansion of the universe with an
equation of state
${\rm P}/\rho =\omega \simeq -1$ as reported recently in \cite{astier}. \\

\centerline{\bf{Acknowledgements}} \vskip .2cm MA and MB
acknowledge CONICET and UNMdP (Argentina) for financial support. \\

\end{document}